\documentclass[12pt]{article}
 \pdfoutput=1
\usepackage{amsfonts,amsthm,amsmath,amssymb,slashed}
\usepackage[textwidth = 500 pt, textheight = 630 pt]{geometry}
\usepackage{latexsym,amsfonts,amsmath,amssymb,theorem,dsfont,wasysym}
\usepackage{amssymb}
\usepackage{amsmath}
\usepackage{amstext}
\usepackage{graphicx,epsfig}
\usepackage{epsfig}
\usepackage{verbatim} 
\usepackage{fancyhdr}
\usepackage{fancybox}
\usepackage{color}
\usepackage{ulem,bbold}
\usepackage{enumitem}
\usepackage{subfigure}
\usepackage{bbm}
\usepackage{parskip}
\usepackage[numbers,sort&compress]{natbib}

\definecolor{MyDarkBlue}{rgb}{0.15,0.15,0.45}
\usepackage[linktocpage=true]{hyperref}
\hypersetup{
colorlinks=true,
citecolor=MyDarkBlue,
linkcolor=MyDarkBlue,
urlcolor=MyDarkBlue,
}

\usepackage[numbers,sort&compress]{natbib}

\def\beq{\begin{eqnarray}}
\def\eeq{\end{eqnarray}}

\def\({\left(}
\def\){\right)}

\newcommand{\be}{\begin{equation}}
\newcommand{\ee}{\end{equation}}

\def\ea{\end{eqnarray}}
\def\ba{\begin{eqnarray}}

\def\beq{\begin{eqnarray}}
\def\eeq{\end{eqnarray}}

\def\({\left(}
\def\){\right)}

\def\lsim{\mathrel{\rlap{\lower3pt\hbox{\hskip0pt$\sim$}}
     \raise1pt\hbox{$<$}}}         
\def\gsim{\mathrel{\rlap{\lower4pt\hbox{\hskip1pt$\sim$}}
     \raise1pt\hbox{$>$}}}         

\def\lsim{\mathrel{\rlap{\lower3pt\hbox{\hskip0pt$\sim$}}
     \raise1pt\hbox{$<$}}}         
\def\gsim{\mathrel{\rlap{\lower4pt\hbox{\hskip1pt$\sim$}}
     \raise1pt\hbox{$>$}}}         

\usepackage{amsmath}
\usepackage{amsfonts}
\usepackage{verbatim}
\usepackage{graphicx}
\usepackage{subfigure}
\usepackage{amssymb}
\usepackage[T1]{fontenc}
\usepackage{slashed}

\begin{document}

\renewcommand{\thefootnote}{\fnsymbol{footnote}}

\makeatletter
\@addtoreset{equation}{section}
\makeatother
\renewcommand{\theequation}{\thesection.\arabic{equation}}

\rightline{}
\rightline{}



\begin{center}
{\Large \bf{Dynamical Friction in Superfluids}}

 \vspace{1truecm}
\thispagestyle{empty} \centerline{\large  {Lasha  Berezhiani $^{1,2}$, Benjamin Elder $^3$ and Justin Khoury $^4$}
}

 \textit{$^1$Max-Planck-Institut f\"ur Physik, F\"ohringer Ring 6, 80805 M\"unchen, Germany\\
 \vskip 5pt
$^2$ Arnold Sommerfeld Center, Ludwig-Maximilians-Universit\"at, \\Theresienstra{\ss}e 37, 80333 M\"unchen, Germany\\
 \vskip 5pt
 $^3$School of Physics and Astronomy, University of Nottingham, \\
 Nottingham, NG7 2RD, United Kingdom\\
 \vskip 5pt
 $^4$Center for Particle Cosmology, Department of Physics and Astronomy, \\ University of Pennsylvania, 209 South 33rd St, Philadelphia, PA 19104, USA
 }

\end{center}  
 
\begin{abstract}
We compute the dynamical friction on a small perturber moving through an inviscid fluid, {\it i.e.}, a superfluid.  Crucially, we account for the tachyonic gravitational mass for sound waves, reminiscent of the Jeans instability of the fluid, which results in non-zero dynamical friction even for subsonic velocities. Moreover, we illustrate that the standard leading order effective theory in the derivative expansion is in general inadequate for analysing supersonic processes. We show this in two ways: (i) with a fluid treatment, where we solve the linearized hydrodynamical equations coupled to Newtonian gravity; and (ii) with a quasiparticle description, where we study the energy dissipation of a moving perturber due to phonon radiation.  Ordinarily a subsonic perturber moving through a superfluid is kinematically prohibited from losing energy, however the Jeans instability modifies the dispersion relation of the fluid which can result in a small but non-vanishing dynamical friction force. We also analyse the soft phonon bremsstrahlung by a subsonic perturber scattered off an external field.

\end{abstract}

\newpage
\setcounter{page}{1}

\renewcommand{\thefootnote}{\arabic{footnote}}
\setcounter{footnote}{0}

\linespread{1.1}
\parskip 4pt

\section{Introduction}
Dynamical friction is a phenomenon wherein an object moving through a cloud of lighter particles experiences a gravitational attraction to its own wake.  The gravitational attraction between the object and cloud results in an overdensity of the cloud behind the object, which slows the motion of the object.

Originally described by Chandrasekhar for a collisionless medium~\cite{Chandrasekhar:1943ys}, this effect has also been examined in gaseous media.  Steady state solutions, where a constant-velocity perturber of mass $M$ moves through an infinite homogeneous gas have shown the dynamical friction to be~\cite{Dokuchaev:1964,Ruderman:1971,Rephaeli:1980,Ostriker:1998fa}
\be
F_\mathrm{DF} = \begin{cases}
0 & V < c_s~; \\
\frac{4 \pi G^2 M^2 \rho_0}{V^2} \ln \frac{R_{\rm SF}}{R_{\rm object}} & V > c_s~,
\end{cases}
\label{DF-nomass}
\ee
where $R_{\rm object}$ and $R_{\rm SF}$ are the sizes of the perturber and homogeneous fluid cloud, respectively. See~\cite{Kim:2007zb,Kim:2008ab,Namouni:2009yc} for more recent work on the subject. 

The traditional approach for obtaining dynamical friction is by means of the hydrodynamic equations together with Poisson's equation:
\beq
\label{conteq}
&&\frac{\partial \rho}{\partial t} + \vec \nabla \cdot (\rho \vec v) = 0\,; \\
\label{eulereq}
&&\frac{\partial \vec v}{ \partial t} + (\vec v \cdot \vec \nabla) \vec v = - \frac{1}{\rho} \vec \nabla P - \vec \nabla \Phi\,;\\
\label{poissoneq}
&&\Delta \Phi = 4 \pi G (\rho+\rho_{\rm pert})\,,
\eeq
where $\Delta$ is the spatial Laplacian, supplemented with the equation of state $P(\rho)$. Here $\rho$, $P$ and $\vec{v}$ are the density, pressure, and velocity of the fluid, respectively, $\Phi$ is the Newtonian potential, and $\rho_{\rm pert}$ represents the density of a perturber. The standard procedure for calculating the dynamical friction is to evaluate the wake configuration left by a moving perturber in the fluid by gravitational interaction and subsequent gravitational pull from this wake acting on the perturber itself. This is somewhat similar to the effect of radiation reaction in electrodynamics, where the electromagnetic field of an accelerated charge acts on the charge itself and slows it down~\cite{Landau:1982dva}.

As it stands, the set of equations~\eqref{conteq}-\eqref{poissoneq} describes a fluid without viscosity, interacting with a perturber gravitationally. Usually, one considers the perfectly inviscid limit by focusing purely on potential flows.  This approach accounts only for longitudinal perturbations, and it ignores transverse modes. For a regular fluid, this is merely an approximation as there always are live transverse degrees of freedom. Classically, this is acceptable because at the level of equation at hand ({\it i.e.}, for a perfect fluid) the energy cost to produce transverse modes vanishes --- their dispersion relation is $\omega_k = 0$.  In other words, production of transverse modes costs no energy and therefore cannot lead to friction.\footnote{Another notable  point, in this context, is Kelvin's circulation theorem, which states that in a barotropic ideal fluid the circulation around a closed curve moving with the fluid is conserved.} Quantum mechanically, this leads to inconsistencies and an extension of Euler's equation is necessary in order to avoid strong coupling~\cite{Endlich:2010hf}. Any such extension would lead to energy dissipation, therefore one needs to be cautious when ascribing the properties of the superfluid to a regular perfect fluid. 

For superfluids at zero temperature, on the other hand, there are indeed only longitudinal dynamical low energy excitations due to absolute degeneracy of the constituents.\footnote{Because of this, the hydrodynamic equation for such a fluid is usually formulated as~\cite{Landau6}
\beq
\frac{\partial \vec{v}}{\partial t}+\vec{\nabla}\left(\frac{1}{2}v^2+\tilde{\mu}+\Phi\right)=0\,,
\eeq
where $\tilde{\mu}$ denotes the chemical potential per unit mass. It must be pointed out that the term involving $\tilde{\mu}$ reduces to the pressure term in Euler's equation upon taking into account that $\rho=\partial P/\partial \tilde{\mu}$. Notice that this equation is in general different from~\eqref{eulereq}.
} Although the above-mentioned hydrodynamic approach to dynamical friction is applicable to superfluids in certain limiting cases, in general it is plagued with puzzles. For instance, as it stands~\eqref{DF-nomass} has an unphysical $c_s\rightarrow 0$ limit. Another puzzling behavior lies in the fact that the supersonic dynamical friction increases with decreasing $V$, until $V = c_s$ is reached, at which point the dynamical friction force discontinuously drops to zero.  One possible resolution for this discontinuity was explored in~\cite{Ostriker:1998fa}, where the perturbing object is only switched on at $t = 0$ and then moves at constant velocity $V$ thereafter.  The time-dependent effects result in a non-zero subsonic force, so that $F_\mathrm{DF}$ is a continuous function of $V$. However, this observation that in this case there is a subsonic drag force (eliminating the discontinuity) does not present a satisfactory resolution of the puzzle. This is because, after a finite amount of time has gone by, the disturbance initiated by the appearance of an object should pass through the entire fluid erasing the memory of the initial event. In other words, one can always find physical situations in which the puzzle persists in the hydrodynamical treatment. 

One of the goals of our work is to show that the inclusion of higher-gradient corrections, which are responsible for the so-called ``quantum pressure'', is essential as one arrives at the wrong expression for the drag force otherwise. As we will see, including this term resolves both of the aforementioned puzzles. As has been demonstrated in~\cite{LB}, the regular (non-gravitational) friction for supersonic motion is sensitive to higher-order corrections to the effective theory of phonons. Here we demonstrate that this holds true in the case of dynamical friction, and that phonon action in the leading-order in the derivative expansion is inadequate for finding the supersonic drag force. For a condensate with negligible self-interactions, the related calculation incorporating the quantum pressure has been performed in~\cite{Hui:2016ltb} in the context of Fuzzy Dark Matter~\cite{Hu:2000ke}.

Another interesting effect incorporated in our discussion is the Jeans' instability, which results in a tachyonic mass for sound waves. It has been previously taken into account for supersonic perturbers~\cite{Ruderman:1971,Rephaeli:1980}, but not subsonic ones. It is well known that when an impurity moves through a superfluid with subsonic velocity, there is no friction (so called ``Landau's criterion''). This is because energy-momentum conservation prohibits the radiation of phonons, so the impurity cannot lose energy to the fluid. This was shown by Landau in the absence of gravity. It relies on the fact that the dispersion relation for a phonon in small-momentum limit reduces to $\omega_k=c_s k$. In other words, the speed of an object must be at least $c_s$ for it to be able to dissipate its energy in sound waves, similar to Cherenkov radiation. The superpower of being frictionless originates from the fact that sound waves are the only low-energy excitations present in the superfluid.

Here, we show that the inclusion of the tachyonic gravitational mass term can lead to (otherwise absent) subsonic dissipation. The physics of this effect is most transparent within the quasiparticle formalism, where energy dissipation is the result of radiating phonons.   In this framework a self-sustained superfluid soliton is formed as a result of the balance between the gravitational attraction and the repulsive pressure, supplemented with quantum pressure. This results in a configuration of finite extent. Short-wavelength sound waves are oblivious to this and propagate as in a homogeneous fluid, however there are modes of sufficiently long wavelength that are sensitive to edge effects. Introducing the aforementioned tachyonic mass term corresponds to the rough account of the latter. In other words, perturbations of a superfluid soliton of wavelength comparable to its size are believed to exhibit near-critical behavior, so that the energy cost for their production can be significantly suppressed compared to the non-gravitational case.

The simplest superfluid is a degenerate state of weakly-interacting bosons in the zero-momentum state. In this work, this is precisely the type of fluid for which we would like to perform a thorough analysis. Although we focus on this simple system, our results are easily generalizable to general superfluids. We begin with a computation of dynamical friction within the quasiparticle formalism. The essence of the method is to start with the low energy effective theory of phonons, gravitationally coupled to a probe particle. Energy dissipation corresponds to the gravity-mediated process of radiating phonons by a probe traveling through the superfluid. 

Moreover, we demonstrate that the results of the quasiparticle formalism can be reproduced using the more traditional hydrodynamical method described above. The correspondence is identical to the calculation of radiation reaction in electrodynamics, wherein an accelerated charge experiences a drag force due to the electromagnetic self-force. Alternatively, one finds the same energy-dissipation rate by evaluating the intensity of the electromagnetic radiation at large distances, see~\cite{Landau:1982dva}.

Our motivation to better understand dynamical friction in superfluids stems from the existence of dark matter models in which particles undergo Bose-Einstein condensation in the central regions of galactic halos~\cite{Goodman:2000tg,Hu:2000ke,Hui:2016ltb,Berezhiani:2015pia,Berezhiani:2015bqa,Berezhiani:2017tth} (see also~\cite{Khoury:2016ehj,Hodson:2016rck,Cai:2017buj,Fan:2016rda,Alexander:2018fjp,Hossenfelder:2018iym,Sharma:2018ydn,Ferreira:2018wup,Berezhiani:2018oxf}). Moreover, there seems to be some tension between the dynamical friction inferred from observations and the one predicted by the cold dark matter paradigm~\cite{Debattista:1997bi,Sellwood:2016,Fornaxold,Tremaine}. One of the possible ways to reconcile the two is the suppression of the drag force by means of dark matter superfluidity~\cite{Chandra}.

\section{Quasiparticle description of the superfluid}
\label{quasiparticle}

We begin by modeling dynamical friction on a perturber moving through a superfluid as a process of energy dissipation due to phonon radiation. This requires an effective theory for phonons, describing the propagation and interactions of any relevant degrees of freedom. Let us start by formulating the microscopic theory of the simplest system capable of forming a superfluid:\footnote{We set $c = \hbar = 1$ and adopt the mostly-plus metric signature.}
\beq
S=\int {\rm d}^4 x \sqrt{-g} \left( \frac{1}{16\pi G}R-|\partial \Psi^2|-m^2 |\Psi|^2-\frac{\lambda}{2}|\Psi|^4 -\frac{1}{2}(\partial \chi)^2-\frac{1}{2}M^2\chi^2\right)\,.
\label{action}
\eeq
This describes a massive self-interacting complex scalar field $\Psi$ and a massive real scalar field $\chi$ minimally coupled to gravity. The field $\Psi$ represents the fluid degree of freedom, and $\chi$ the perturber.

Upon condensation, $\Psi$ particles form a superfluid. In the absence of gravity, the superfluid state is described by the following homogeneous classical field configuration
\beq
\Psi=v\,{\rm e}^{{\rm i} \sqrt{m^2+\lambda v^2}t}\,,
\label{class}
\eeq
where $v$ is fixed in terms of the particle number density via $n=2 v^2 \sqrt{m^2+\lambda v^2}$. To derive the effective theory for fluctuations, we perturb the classical background as
\beq
\Psi=(v+h){\rm e}^{{\rm i} \sqrt{m^2+\lambda v^2}t+{\rm i}\pi}\,.
\label{pert}
\eeq
Since we are interested in a gravitating superfluid, we must keep track of gravitational degrees of freedom.  For our purposes it suffices to treat the fluid non-relativistically, hence the only relevant component of the gravitational sector is the Newtonian potential $\Phi$. After substituting~\eqref{pert} back into~\eqref{action}, and linearizing gravity by retaining only leading order terms in $\Phi\ll 1$, we obtain
\ba
\mathcal{L}&=& \frac{1}{8\pi G}\Phi \Delta \Phi- \big(\vec{\nabla}h\big)^2+2m(v+h)^2\left( \mu-m\Phi+\dot{\pi}-\frac{(\vec{\nabla}\pi)^2}{2m}\right)-\frac{\lambda}{2}(v+h)^4\nonumber \\
&& +~\frac{1}{2}\dot{\chi}^2-\frac{1}{2}(\vec{\nabla}\chi )^2-\frac{1}{2}M^2\chi^2-\Phi M^2\chi^2\,.
\label{EFT0}
\ea
In the last term we have assumed that the perturber is non-relativistic, {\it i.e.}, $\dot{\chi}\approx M\chi$,  and $\mu\equiv \frac{\lambda v^2}{2m}$ is the (non-relativistic) chemical potential. Notice that $\mu\ll m$ has been assumed in deriving \eqref{EFT0}, to ensure subluminal sound speed. The next step is to integrate out $h$ using its equation of motion. We will do this to leading order in the derivative expansion, neglecting the $\big(\vec{\nabla}h\big)^2$ term. Later on, in Sec.~\ref{higher grad}, we will come back and refine the result by including higher-gradient terms. The result, to leading order in gradients, is
\beq
\mathcal{L}=\frac{1}{8\pi G}\Phi \Delta \Phi+\frac{2m^2}{\lambda}\left( \mu-m\Phi+\dot{\pi}-\frac{(\vec{\nabla}\pi)^2}{2m}\right)^2 +\frac{1}{2}\dot{\chi}^2-\frac{1}{2}(\vec{\nabla}\chi )^2-\frac{1}{2}M^2\chi^2-\Phi M^2\chi^2\,.
\label{EFT leading}
\eeq
The resulting Lagrangian, not surprisingly, looks like a particular case of the more general effective field theory of a non-relativistic superfluid minimally coupled to gravity~\cite{Son:2005rv}, together with a spectator $\chi$ field also coupled to gravity.\footnote{Here, we have made a non-relativistic approximation for the interaction term between $\chi$ and $\Phi$, but have retained the relativistic form for the $\chi$ kinetic term. This is done purely for aesthetic reasons, as we will use non-relativistic dispersion relation whenever necessary.}

Expanding~\eqref{EFT leading} and rescaling the Goldstone boson via $\pi\rightarrow \frac{\sqrt{\lambda}}{2m}\,\pi $, we arrive at the following action
\ba
\mathcal{L}&=&\frac{1}{8\pi G}\Phi \Delta \Phi +\frac{\rho_0}{2c_s^2}\Phi^2 -\frac{\sqrt{\rho_0}}{c_s} \Phi\dot{\pi}+\frac{1}{2} \Phi (\vec{\nabla} \pi)^2 -\Phi M^2\chi^2\nonumber\\
&&+~\frac{1}{2}\left( \dot{\pi}^2-c_s^2 (\vec{\nabla}\pi)^2 \right)    -\frac{c_s}{2\sqrt{\rho_0}}\dot{\pi} (\vec{\nabla} \pi)^2+\frac{1}{8} \frac{c_s^2}{\rho_0}(\vec{\nabla} \pi)^4 + \frac{1}{2}\dot{\chi}^2-\frac{1}{2}(\vec{\nabla}\chi )^2-\frac{1}{2}M^2\chi^2\,,
\label{withphi}
\ea
where $\rho_0 = mn$ is the mass density of the unperturbed superfluid, and
\beq
c_s^2\equiv \frac{\lambda\rho_0}{4m^4}
\label{sound speed}
\eeq
defines the sound speed. Notice that \eqref{EFT leading} contains a tadpole in $\Phi$, due to the fact that the presence of the fluid sources the gravitational potential. In deriving \eqref{withphi}, such a nontrivial background has been subtracted from $\Phi$ and reabsorbed in $\mu$. Of course in general this would lead to spatially inhomogeneous spectrum of perturbations. Nevertheless, it can be shown \cite{Berezhiani:2015bqa} that within the core of the superfluid soliton (which is the subject of our interest) the quantity $\mu-m\Phi$ assumes a relatively homogeneous background.

Equation~\eqref{withphi} could be used to calculate the process of radiating phonons by moving $\chi$, with~$\Phi$ acting as the mediator. However, it is more convenient to integrate out the Newtonian potential using its equation of motion. Since in this work we are interested only in the energy dissipation process $\chi\rightarrow\chi+\pi$, it suffices to focus on this single interaction vertex generated by integrating out $\Phi$. Thus for our purposes the relevant Lagrangian is 
\beq
\mathcal{L}_\pi=\frac{1}{2}\dot{\pi} \frac{c_s^2 \Delta}{c_s^2\Delta+m_g^2} \dot{\pi}+\frac{1}{2} c_s^2 \pi \Delta \pi
-4\pi G\sqrt{\rho_0}c_s
M^2 \chi^2 \frac{1}{c_s^2\Delta+m_g^2}\dot{\pi}+\ldots\,,
\label{EFT1}
\eeq
where
\beq
m_g^2\equiv 4\pi G\rho_0
\label{Jeans scale}
\eeq
is the tachyonic mass responsible for the Jeans instability, and ellipses denote the free part of the $\chi$ action as well as higher-order terms in $\pi$ and $\chi$. All the information about the microscopic theory~\eqref{action} is now encoded in $c_s$, in particular through its dependence on the couplings and density of the condensate. In other words, for an arbitrary superfluid the relevant terms for radiating a single phonon by a moving perturber would still be given by~\eqref{EFT1}. To bring the action to a more familiar form, it is convenient to make the field redefinition 
\beq
\pi\rightarrow \sqrt{\frac{c_s^2\Delta+m_g^2}{c_s^2\Delta }}\pi\,.
\eeq
As a result, the quadratic action simplifies to
\beq
\mathcal{L}^{(2)}=\frac{1}{2}\dot{\pi}^2+\frac{1}{2}\pi\left( m_g^2 + c_s^2\Delta \right)\pi\,,
\eeq
which, in Fourier space, leads to the familiar dispersion relation
\beq
\omega_k^2=-m_g^2+c_s^2k^2\,.
\label{omegak0}
\eeq
The tachyonic mass term is the gravitational contribution, which is responsible for the Jeans instability.

\subsection{Including higher-gradient corrections}
\label{higher grad}

The effective theory of superfluids, derived above to leading order in the derivative expansion, is valid for processes with momentum transfer $k\ll mc_s$. In general, while still within the non-relativistic approximation ($k\ll m$, $c_s\ll 1$), one may be interested in processes that lie beyond such leading-order description. A perfect example of a such process is supersonic friction in superfluids~\cite{LB}. Here we are interested in dynamical friction due to gravitational interactions. For this, we re-derive the relevant part of the effective theory accounting for all non-relativistic higher-derivative corrections, thereby extending the results of~\cite{LB} for gravitational interactions. 

The starting point, as before, is the non-relativistic action~\eqref{EFT0} for perturbations around the superfluid configuration. The only assumption at this point is the non-relativistic approximation, with $k\ll m$ and $c_s\ll 1$. We proceed once again by integrating out $h$ using its equation of motion, this time taking into account the gradient term $\big(\vec{\nabla}h\big)^2$. Unlike the previous derivation, where $h$ entered the Lagrangian algebraically, the integration procedure must now be done order by order in perturbations. The resulting action is quite cumbersome and we will not present it here. Since we are interested in the process of phonon radiation, $\chi\rightarrow \chi+\pi$, it suffices to derive the higher-derivative extension of \eqref{EFT1}. 

To the leading order in perturbations, $h$ gets expressed in terms of other fields as
\beq
h=\frac{2mv}{-\Delta+4m^2c_s^2}\left(\dot{\pi}-m\Phi \right)\,.
\eeq
Substituting this back in the action, we get
\beq
\mathcal{L}=\frac{1}{8\pi G}\Phi \Delta \Phi-\frac{\rho_0}{2m^2}\big(\vec{\nabla}\pi \big)^2+2\rho_0\left(\dot{\pi}-m\Phi \right)\frac{1}{-\Delta+4 m^2 c_s^2}\left(\dot{\pi}-m\Phi \right)-\Phi M^2\chi^2+\ldots \,,
\eeq
where the ellipses stand for higher-order terms in perturbations, irrelevant for our discussion, and for the free part of the $\chi$-action.  
Integrating out the Newtonian potential and rescaling $\pi\rightarrow \frac{mc_s}{\sqrt{\rho_0}}\, \pi $ for convenience, we obtain
\ba
\mathcal{L} &=&\frac{1}{2}\dot{\pi} \frac{c_s^2 \Delta}{c_s^2\Delta-\frac{\Delta^2}{4m^2}+m_g^2} \dot{\pi}+\frac{1}{2} c_s^2 \pi \Delta \pi-4\pi G M^2\sqrt{\rho_0}c_s \chi^2 \frac{1}{c_s^2\Delta-\frac{\Delta^2}{4m^2}+m_g}\dot{\pi} \nonumber \\
&&-~2\pi G M^4 \chi^2 \frac{c_s^2-\frac{\Delta}{4m^2}}{c_s^2\Delta-\frac{\Delta^2}{4m^2}+m_g^2} \chi^2 +\ldots
\label{EFT nonlocal}
\eeq
The second line describes a non-local self-interaction terms for $\chi$, arising from the integration of $\Phi$. In the limit $m_g\rightarrow 0$, this term reduces to the Newtonian interaction between two $\chi$-sources. This limit is applicable for processes deep inside the superfluid core. If we were to take the separation between sources to be comparable to Jeans scale $k_{\rm J}^{-1}$, then the graviton-phonon mixing would seem to modify the long range interaction in an interesting way. To make any quantitative statement about this modification, one would need to study the spectrum of the finite superfluid core more precisely, which is beyond the scope of this paper. 

In order to calculate the dissipation rate using standard scattering relations, we perform the canonical normalization by
\beq
\pi\rightarrow \sqrt{\frac{m_g^2+c_s^2\Delta-\frac{\Delta^2}{4m^2}}{c_s^2 \Delta}}\, \pi\,.
\eeq
(The argument of the square root is positive for large enough $k$, as will be made precise later.)
The first line of~\eqref{EFT nonlocal} becomes
\beq
\mathcal{L}=\frac{1}{2}\dot{\pi}^2+\frac{1}{2} \pi \left( m_g^2+c_s^2\Delta-\frac{\Delta^2}{4m^2} \right) \pi-4\pi GM^2\sqrt{\rho_0} \chi^2  \frac{1}{\sqrt{\Delta \left( m_g+c_s^2\Delta-\frac{\Delta^2}{4m^2} \right)}}\, \dot{\pi}\,.
\label{highderL}
\eeq
We can easily read off the dispersion relation for phonons,
\beq
\omega_k^2=-m_g^2 +c_s^2k^2+\frac{k^4}{4 m^2}\,.
\label{highderdisp}
\eeq
This generalizes~\eqref{omegak0} to include a higher-gradient correction. 

\subsection{Free Motion: Phonon Radiation}

Consider a freely-moving particle $\chi$ with uniform velocity $V$. The process of interest is $\chi\rightarrow \chi+\pi$, in which the $\chi$ particle with initial energy-momentum $p_\chi^{{\rm in}}=(E_\chi^{{\rm in}},\vec{p}_{\chi}^{~{\rm in}})$ radiates a phonon with $p_\pi^f=(\omega_k,\vec{k})$. The rate of such process can be straightforwardly computed using
\beq
{\rm d}\Gamma=\frac{1}{2E_{\chi}^{{\rm in}}} \,\frac{{\rm d}^3p_\chi^f}{(2\pi)^32E_\chi^f}\,\frac{{\rm d}^3k }{(2\pi)^32\omega_k} \, |\mathcal{A}|^2(2\pi)^4 \delta^{(4)}(p_\chi^{{\rm in}}-p_\chi^{f}-p_\pi^{f})\,.
\label{rate 1}
\eeq
The radiation amplitude $\mathcal{A}$ readily follows from~\eqref{highderL},
\beq
\mathcal{A}=8\pi GM^2\sqrt{\rho_0} \frac{\omega_k}{\sqrt{k^2\left( -m_g^2 +c_s^2k^2+\frac{k^4}{4 m^2} \right)}}=8\pi GM^2 \frac{\sqrt{\rho_0}}{k}\,.
\eeq
In the last step we have used~\eqref{highderdisp}, which is allowed since we are interested in the amplitude for on-shell phonon radiation.
From~\eqref{rate 1} we may easily obtain the energy dissipation rate, and therefore the friction force
\beq
|F|=\frac{\dot{E}}{V}=\int\omega_k\,{\rm d}\Gamma\,.
\eeq

Since we are interested in the leading-order result in the non-relativistic limit, the momenta can be parameterized as
\beq
&&p_\chi^{{\rm in}}=\left(M+\frac{MV^2}{2},M\vec{V}\right)\,,\\
&&p_\chi^f=\left(M+\frac{MU^2}{2}, M\vec{U}\right)\,,\\
&&p_\pi^f=\left(\omega_k, \vec{k}\right)\,.
\eeq
After performing the integral over the final momentum of $\chi$, we obtain the following expression for the friction force
\beq
|F|=\frac{4\pi G^2M^2\rho_0}{V^2} \int_{k_{\rm min}}^{k_{\rm max}} {\rm d}{\rm cos }~\theta~ \frac{{\rm d}k}{k}\,\delta\left( {\rm cos }~\theta-\frac{\frac{k^2}{2M}+\omega_k}{kV}\right)\,.
\label{Fdelta}
\eeq
In deriving this expression, we have used the fact that both initial and final $\chi$ are non-relativistic and have replaced the corresponding energies in the expression of the rate with the mass gap.  

The limits of integration, $k_{\rm min}$ and $k_{\rm max}$, set the range of scales over which the effective description of the superfluid is valid:

\begin{itemize}

\item {\bf High-momentum cutoff, $k_{\rm max}$}: Since we are working in the non-relativistic limit, the superfluid description (in particular~\eqref{highderdisp}) breaks down at $k\simeq m$. Therefore, $m$ is an obvious candidate for $k_{\rm max}$.  Although it suffices to describe the perturber as point-like, if we wanted to generalize the result to an extended classical object like a star then we should instead cut the integration off at the momentum scale set by the size of the extended perturber, $k_{\max} \approx R_{\rm object}^{-1}$. This is because if we were to consider modes shorter than the size of the object then we would also have to consider the internal degrees of freedom of the perturber. In other words, in general the high-momentum cut-off sets the scale at which ignored degrees of freedom become dynamical, whether these degrees of freedom belong to the fluid or the moving perturber.

The angular integration in~\eqref{Fdelta} is easy to perform using the delta-function. All we need to do is to ensure that the value of the cosine we get from delta-function is less than unity, which can be incorporated as an integration limit for the momentum integral. In other words, the condition $\cos \theta<1$ gets translated into $k<k_\star$, with $k_\star$ defined by
\beq
\frac{k_\star^2}{2M}+\omega_{k_\star}=k_\star V\,.
\label{kstar}
\eeq
This sets the upper bound of the momentum integral:
\be
k_{\max} = \min \left( 2 \pi R_{\rm object}^{-1}, k_\star \right)\,.
\label{kmax def}
\ee

\item  {\bf Low-momentum cutoff, $k_{\rm min}$}: The low-momentum cut-off $k_{\rm min}$ is set by the length scale over which the assumed homogeneity of the superfluid ceases to be a good approximation. For instance, it could be set by the spatial extent of the superfluid phase, $k_{\min} \approx R_{\rm SF}^{-1}$. On the other hand, the longest possible wavelength for which the description can be trusted is given by the Jeans scale, $k_{\rm J}$. From the phonon dispersion relation~\eqref{highderdisp}, the Jeans scale is given by 
\beq
k_{\rm J}^2=2m^2c_s^2\left( -1+\sqrt{1+\frac{m_g^2}{m^2c_s^4}} \right)\,.
\label{jeansk}
\eeq
Any mode softer than $k_{\rm J}$ would exhibit tachyonic behavior. This is reminiscent of the fact that gravitational instability tends to destroy the coherence over scales greater than the Jeans wavelength. In general, therefore, the large-distance cutoff is set by either the Jeans scale or the size of the superfluid cloud, whichever is smaller. This implies the low-momentum cutoff
\beq
k_{\min} = \max \left( 2 \pi R_{\rm SF}^{-1}, k_\mathrm{J} \right) \,.
\label{kmin def}
\eeq

\end{itemize}

After performing the angular integral, as described above, we are left with the following expression for the friction force,
\beq
|F| =\frac{4\pi G^2M^2\rho_0}{V^2} \int_{k_{\min}}^{k_{\max}}  \frac{{\rm d}k}{k} = \frac{4\pi G^2M^2\rho_0}{V^2}~{\rm ln}\left( \frac{ \min \left( 2\pi R_{\rm object}^{-1}, k_\star\right)}{ \max \left( 2 \pi R_{\rm SF}^{-1}, k_\mathrm{J} \right)} \right)\,.
\label{Fkint}
\eeq
It must be stressed that the integral is only non-vanishing provided that $k_{\max} > k_{\min}$. We will study various limits of this expression in Sec.~\ref{limits sec}.
Before doing that, it is instructive to rederive this result using a hydrodynamical description.

\section{Hydrodynamical description}

In this Section we adopt the standard hydrodynamical approach to computing dynamical friction for a superfluid. Our setup closely follows that of~\cite{Ostriker:1998fa}.  A key difference is that we include the Jeans instability of the fluid, which results in a non-vanishing subsonic dynamical friction force.\footnote{It should be noted that~\cite{Ruderman:1971,Rephaeli:1980} considered steady-state supersonic motion with the Jeans instability, as we do here.  However, neither paper computed the same effect for subsonic motion.}
Since we are interested in superfluids, we also keep track of the ``quantum pressure'' term defined below. A similar calculation has been performed by \cite{Kahn} for $c_s = 0$, ignoring the effects due to the Jeans instability.

The starting point is the action~\eqref{action} for the complex scalar field $\Psi$. To take the non-relativistic limit, we rescale the field as follows,
\be
\Psi = \frac{\psi}{\sqrt{2m}} \,{\rm e}^{-{\rm i} mt}\,,
\ee
assume $|\partial_t \psi | \ll | m \psi |$, and work in the Newtonian approximation for the gravitational field. The resulting non-relativistic Lagrangian density for the fluid coupled to gravity is
\be
{\cal L} = \frac{1}{8\pi G}\Phi \Delta \Phi + \frac{{\rm i}}{2} \left(\psi^\star \partial_t\psi - \psi \partial_t\psi^\star\right) - \frac{|\vec{\nabla} \psi|^2}{2m} - \frac{\lambda}{8m^2} |\psi|^4 - m|\psi|^2 \Phi\,.
\ee
The equation of motion for $\psi$ is the well-known Gross-Pitaevskii equation
\be
{\rm i}\partial_t\psi = - \frac{\Delta}{2m} \psi + \frac{\lambda}{4m^2}|\psi|^2 \psi  + m\Phi\psi\,.
\label{Schrod}
\ee
To cast this equation in hydrodynamical form we follow~\cite{Boehmer:2007um} and decompose the wavefunction as
\be
\psi = \sqrt{\frac{\rho}{m}} {\rm e}^{{\rm i}\theta}\,,
\ee
where $\rho$ is identified as the fluid density, $\vec{v} = \frac{\vec{\nabla}\theta}{m}$ defines the fluid velocity, and $P = \frac{\lambda}{8m^4}\rho^2$ is the pressure. 
The real and imaginary parts of~\eqref{Schrod} give the mass conservation and Euler equations:
\ba \nonumber
\frac{\partial \rho}{\partial t} + \vec \nabla \cdot (\rho \vec v) &=& 0\,; \\
\frac{\partial \vec v}{ \partial t} + (\vec v \cdot \vec \nabla) \vec v &=& - \frac{1}{\rho} \vec \nabla P - \vec \nabla \Phi + \frac{1}{2 m^2} \vec \nabla \left( \frac{\Delta \sqrt \rho}{\sqrt \rho} \right)\,.
\label{navier-stokes}
\ea
The last term in the second equation is the so-called ``quantum pressure'' term. It is important to note that although~\eqref{navier-stokes} was derived for the quartic potential, the resulting equations hold more generally for any equation of state $P = P(\rho)$.

\subsection{Linearized fluid equations}

To proceed, we perturb the hydrodynamical equations around a static background of density $\rho_0$,
\be
\rho = \rho_0(1 + \alpha)\,.
\ee
We assume that $\rho_0$ is approximately homogeneous, which is valid on scales smaller than the Jeans length, as we will make precise later.  
To linear order the equations become
\ba
\nonumber
& & \dot{\alpha} + \vec{\nabla}\cdot \vec{v} = 0\,; \\
& & \dot{\vec{v}} + c_s^2\vec{\nabla} \alpha = - \vec \nabla \phi + \frac{\Delta}{4m^2} \vec{\nabla}\alpha\,,
\label{hydro linear}
\ea
where $c_s^2 = \frac{\partial P}{\partial \rho}\big|_{\rho_0}$ is the fluid sound speed, and $\phi$ is the perturbation in the gravitational potential. 
The latter is sourced by the fluid overdensity as well as the perturber $\rho_\mathrm{ext}$,
\be
\Delta \phi = 4 \pi G\left(\rho_\mathrm{ext} + \rho_0 \alpha\right)\,.
\label{poisson-eqn}
\ee
Equations~\eqref{hydro linear} and~\eqref{poisson-eqn} can be combined as usual to yield a wave equation for the overdensity:
\be
\ddot{\alpha} - c_s^2 \Delta\alpha -m_g^2 \alpha + \frac{1}{4m^2} \Delta^2\alpha = 4 \pi G \rho_\mathrm{ext}\,,
\label{wave-eqn}
\ee
where $m_g$ is the tachyonic mass scale defined in~\eqref{Jeans scale}. The inclusion of the fluid overdensity in the Poisson equation~\eqref{poisson-eqn} is a key difference between our approach and that of~\cite{Ostriker:1998fa}, which only included the perturber. The overdensity contribution gives rise to the tachyonic mass term in~\eqref{wave-eqn}, which is responsible in turn for the Jeans instability.

\subsection{Dynamical friction on perturber}

The perturber is described by a particle of mass $M$ moving at constant velocity $V$, which is assumed without loss of generality
to point in the $z$ direction: 
\be
\rho_\mathrm{ext}(x) = M \delta(x) \delta(y) \delta(z - Vt)\,.
\ee
The perturber sources a fluid overdensity via~\eqref{wave-eqn}, which in turn generates a gravitational force on the particle.

The force acting on the perturber is proportional to the rate of change of the energy.  Since the velocity of the perturber is assumed to be constant,
only the change in potential energy needs to be considered:
\be
F = \frac{M}{V} \dot \phi_\alpha \,,
\ee
where $\phi_\alpha$ is that part of the gravitational potential sourced by the linear fluid overdensity: 
\be
\phi_\alpha(k_0,\vec{k}) = - \frac{4 \pi G \rho_0}{\vec{k}^2}\alpha(k_0,\vec{k}) \,.
\ee
Meanwhile, the fluid overdensity is determined by~\eqref{wave-eqn}:
\be
\alpha(k_0,\vec{k}) = - \frac{4 \pi G \rho_\mathrm{ext}(k)}{ k_0^2 - \omega_k^2} \,,
\ee
where $\omega_k$ satisfies the dispersion relation~\eqref{highderdisp}. Putting everything together, we obtain\footnote{Equivalently, we could have computed the force as the gradient of the gravitational potential $\vec{F}=- M \vec \nabla \Phi$. In fact, it is easy to see from the last line of \eqref{DF-moment} that the gradient can be traded for the time derivative due to the appearance of delta function.}
\ba
\nonumber
F &=& \frac{M}{V} \int \frac{\mathrm{d}^4 k}{(2 \pi)^4} \, {\rm e}^{{\rm i} k_0 t - {\rm i} \vec k \cdot \vec x} \, {\rm i} k_0\, \frac{4 \pi G \rho_0}{\vec{k}^2}\, \frac{4 \pi G \rho_\mathrm{ext}(k)}{k_0^2 - \omega_k^2} \\
&=& \frac{2 {\rm i}}{\pi} \frac{G^2 M^2 \rho_0}{V} \int {\rm d}^4 k\, {\rm e}^{{\rm i} k_0 t - {\rm i} \vec k \cdot \vec x} \,\frac{k_0}{\vec{k}^2} \,\frac{\delta(k_0 - k_z V)}{k_0^2 - \omega_k^2}\,,
\label{DF-moment}
\ea 
We restrict our attention to modes with sufficiently large wavenumber such that $\omega_k$ is real.
That is, we only integrate over modes with wavelength smaller than the Jeans scale, $k > k_{\rm J}$, where
\be
k_{\rm J}^2 \equiv 2 m^2 c_s^2 \left( -1 + \sqrt{1 + \frac{m_g^2}{c_s^4 m^2}} \right)\,.
\label{kjeans}
\ee

The $k_0$ integral is done via contour integration, closing the contour from above and below for $t > 0$ and $t < 0$, respectively. The poles at $k_0 = \pm \omega_k$ must be pushed infinitesimally off the real axis. Since the force must be non-zero at all times, we push one pole up and the other pole down. The answer turns out to be independent of which pole is pushed up versus down, hence we will assume the prescription $k_0 = \pm\left(\omega_k + {\rm i}\epsilon\right)$. The result for $t > 0$ is
\be
F = -\frac{2G^2 M^2 \rho_0}{V} \int {\rm d}^3 k \frac{ {\rm e}^{{\rm i}\omega_k t - {\rm i}\vec k \cdot \vec x}}{\vec{k}^2} \delta(\omega_k - k_z V) \,. 
\ee
The force must be evaluated at the location of the perturber, $\vec{x} = Vt\hat{z}$. Using spherical coordinates, the azimuthal integral is trivial, leaving us with
\be
F = -\frac{4 \pi G^2 M^2 \rho_0}{V^2} \int \frac{{\rm d}k}{k}\, {\rm d}\cos\theta\, {\rm e}^{{\rm i} \omega_k t - {\rm i} k Vt \cos \theta} \delta \left( \frac{\omega_k}{k V} - \cos \theta \right) \,.
\label{DF-kcos}
\ee
In order for a particular wavenumber $k$ to contribute to the dynamical friction force, the argument of the $\delta$-function~\eqref{DF-kcos} must be non-zero for some $\theta$. 
This is the case for $k \leq k_\star$, where $k_\star$ is determined by
\be
\omega_{k_\star} = k_\star V\,.
\label{kstar-def}
\ee
This agrees with~\eqref{kstar} in the $M\rightarrow \infty$ limit. This is so because~\eqref{kstar-def} does not account for the recoil of the perturber,
as it is assumed to be moving with a constant velocity. 

The range of $k$ is further constrained by the regime of validity of our analysis. Clearly the analysis is only valid for scales smaller than the size $R_{\rm SF}$ of the homogeneous gas cloud, thus $k \geq \frac{2 \pi}{R_{\rm SF}}$. Similarly, we do not consider modes with wavelength smaller than $R_{\rm object}$, where $R_{\rm object}$ is either the physical size of the perturber or is the distance below which the fluid linearization breaks down and $\alpha \gsim 1$. Putting these requirements together, we only integrate over a particular range of momenta, if it exists at all:
\be
k_{\min} \leq k \leq k_{\max}\,,
\ee
where
\be
k_{\min} = \max \big( 2 \pi R_{\rm SF}^{-1}, k_\mathrm{J} \big) \,; \qquad  k_{\max} = \min \big(2 \pi R_{\rm object}^{-1} , k_\star \big)\,.
\ee

We can now do the angular and $k$ integrals in Eq.~\eqref{DF-kcos} to obtain
\be
\left| F \right| = \frac{4 \pi G^2 M^2 \rho_0}{V^2} \ln \left( \frac{ \min \big( 2 \pi R_{\rm object}^{-1} , k_\star \big) }{ \max \big( 2 \pi R_{\rm SF}^{-1}, k_\mathrm{J}  \big) } \right)\,.
\label{ForceDF}
\ee
This result is identical to~\eqref{Fkint}, obtained using the quasiparticle description. The only difference is that the quasiparticle derivation accounted for the change in momentum of the perturber, resulting in a slightly more general expression for $k_\star$. In the limit of very heavy perturber, $M \to \infty$, the two results agree exactly.

To complete the calculation, we are now in a position to check the validity of approximating $\rho_0$ as homogeneous. For simplicity we will ignore the quantum pressure term. In this approximation, the background profile is determined by hydrostatic equilibrium: $\vec \nabla P_0 = -\rho_0\vec \nabla \Phi$.  Substituting our equation of state $P_0 = \frac{\lambda}{8m^4} \rho^2_0$, and taking the divergence on both sides of the equation, we obtain
\be
\Delta\rho_0 = - \frac{16\pi G m^4}{\lambda}\rho_0\,,
\ee
where we have used the background Poisson equation. This is the well-known Lane-Emden equation for $P \sim \rho^2$.
The resulting density profile is a nearly constant density core which terminates at~\cite{Goodman:2000tg,Boehmer:2007um} 
\be
R_{\rm SF} = \pi \sqrt{\frac{\lambda}{16\pi G m^4}} \,.
\ee
This agrees with the Jeans scale, which in this approximation is
\be
k_{\rm J}^{-1} \simeq \frac{c_s}{m_g} = \sqrt{\frac{\lambda}{16\pi G m^4}}\,,
\ee
where we have used~\eqref{sound speed}. Thus the approximation of nearly uniform $\rho_0$ is valid on scales much smaller than the Jeans' scale, as claimed at the onset.

\section{Limiting cases}
\label{limits sec}

The general expression~\eqref{Fkint} for the dynamical friction force involves the somewhat cumbersome expressions for $k_\star$ and $k_\mathrm{J}$. To gain intuition, in this Section we explore different limits in which the expressions simplify considerably.

\subsection{Small superfluid core (compared to the Jeans scale)}

Consider the limit where the size of the homogeneous superfluid cloud is considerably smaller than the Jeans scale, {\it i.e.}, $R_{\rm SF} \ll k_{\rm J}^{-1}$. In other words, one can think of this regime as the $k_{\rm J}\rightarrow 0$ limit. By definition, it follows from~\eqref{kmin def} that
\be
k_{\min} = 2 \pi R_{\rm SF}^{-1}\,.
\ee
For simplicity, we consider the case of a sufficiently localized perturber, such that $R_{\rm object}^{-1} \ll k_\star$. As we will see, this only affects the result in the highly-supersonic regime, $V\gg c_s$. With this assumption, it follows from~\eqref{kmax def} that 
\be
k_{\max} = k_\star\,.
\ee
To determine $k_\star$, we must solve~\eqref{kstar}. Ignoring the tachyonic mass in the dispersion relation~\eqref{highderdisp}, such that $\omega_k^2 \simeq c_s^2k^2+\frac{k^4}{4 m^2}$,~\eqref{kstar} reduces to 
\be
\frac{k_\star}{2M}+\sqrt{c_s^2+\frac{k^2_\star}{4 m^2}}= V\,.
\label{kstar simple}
\ee
Manifestly, for $V < c_s$ there is no real and positive solution. Correspondingly, because of the delta function in~\eqref{Fdelta}, the force vanishes for subsonic motion:
\be
|F| = 0\,;\qquad V < c_s\,.
\label{F=0}
\ee
For $V>c_s$, on the other hand, the solution to~\eqref{kstar simple} is the more palatable expression:
\beq
k_\star=\frac{2Mm^2}{M^2-m^2}\left( - V+\sqrt{\frac{M^2}{m^2}(V^2-c_s^2)+ c_s^2} \right)\,;\qquad V > c_s\,.
\label{kstar simple 2}
\eeq
An interesting point that is easily checked is that $k_\star\ll m$ as long as~$V\ll 1$. This means that, in the case of a non-relativistic perturber, the non-relativisitic field theory formalism is justified for calculating the drag force for arbitrarily small $m$. Equation~\eqref{kstar simple 2} simplifies further in the limit~$M\rightarrow \infty$, 
\be
\lim_{M\rightarrow \infty}k_\star=2mc_s\sqrt{\frac{V^2}{c_s^2}-1}\,;\qquad V > c_s\,.
\ee
This regime of a heavy perturber is physically justified in the astrophysical context of interest, where $M$ is the largest mass scale at hand.

The drag force~\eqref{Fkint} is non-zero provided that $k_{\rm max} = 2mc_s\sqrt{\frac{V^2}{c_s^2}-1}$ is larger than $k_{\rm min} = 2 \pi R_{\rm SF}^{-1}$,
that is, for $V \geq c_s \sqrt{1 + \frac{\pi^2}{m^2c_s^2R_{\rm SF}^2}}$. The result is
\beq
|F|
=\left\{\begin{array}{cl}
\frac{4\pi G^2M^2\rho_0}{V^2}~{\rm ln}\left( \frac{mc_sR_{\rm SF}}{\pi} \sqrt{\frac{V^2}{c_s^2}-1} \right)  \hspace{20pt}&\text{for}\hspace{10pt} V \geq c_s \sqrt{1 + \frac{\pi^2}{m^2c_s^2R_{\rm SF}^2}}\,; \\ \\
0 \hspace{20pt}&\text{for}\hspace{10pt} V \leq c_s \sqrt{1 + \frac{\pi^2}{m^2c_s^2R_{\rm SF}^2}} \,.
\end{array}\right.
\label{DF-qp}
\eeq
Thus the friction force vanishes smoothly as we approach the sonic limit, albeit at a finite difference from $c_s$. The distinction between our result and the one given by~\eqref{DF-nomass} is obvious and arises from including higher-gradient corrections in the dispersion relation. Note that had we kept $R_{\rm object}$ around, we would find that the force is modified for $ V^2  > c_s^2+\frac{\pi^2}{R_{\rm object}^2 m^2}$, resulting instead in $|F| \simeq \frac{4\pi G^2M^2\rho_0}{V^2}~{\rm ln}\frac{R_{\rm SF}}{R_{\rm object}}$. Equation~\eqref{DF-qp} is plotted in the Left Panel of Fig.~\ref{DF-plot} for different values of $mc_sR_{\rm SF}$.

\subsection{\bf Superfluid core of critical size ({\it i.e.}, of the size of the Jeans scale):} 

Consider the case where the homogeneous superfluid cloud is of comparable size to the Jeans scale, {\it i.e.}, $R_{\rm SF} \sim  k_{\rm J}^{-1}$. It follows trivially from~\eqref{kmin def} that the IR cutoff for the momentum integrals is
\be
k_{\min} \simeq k_{\rm J}\,.
\ee
This implies there are macroscopic superfluid excitations with vanishing energy cost. Therefore one would expect new effects not captured by the limit considered earlier. 
One potentially interesting regime, as we will see, is subsonic dissipation.

\begin{figure}
\includegraphics[width = 0.5\linewidth]{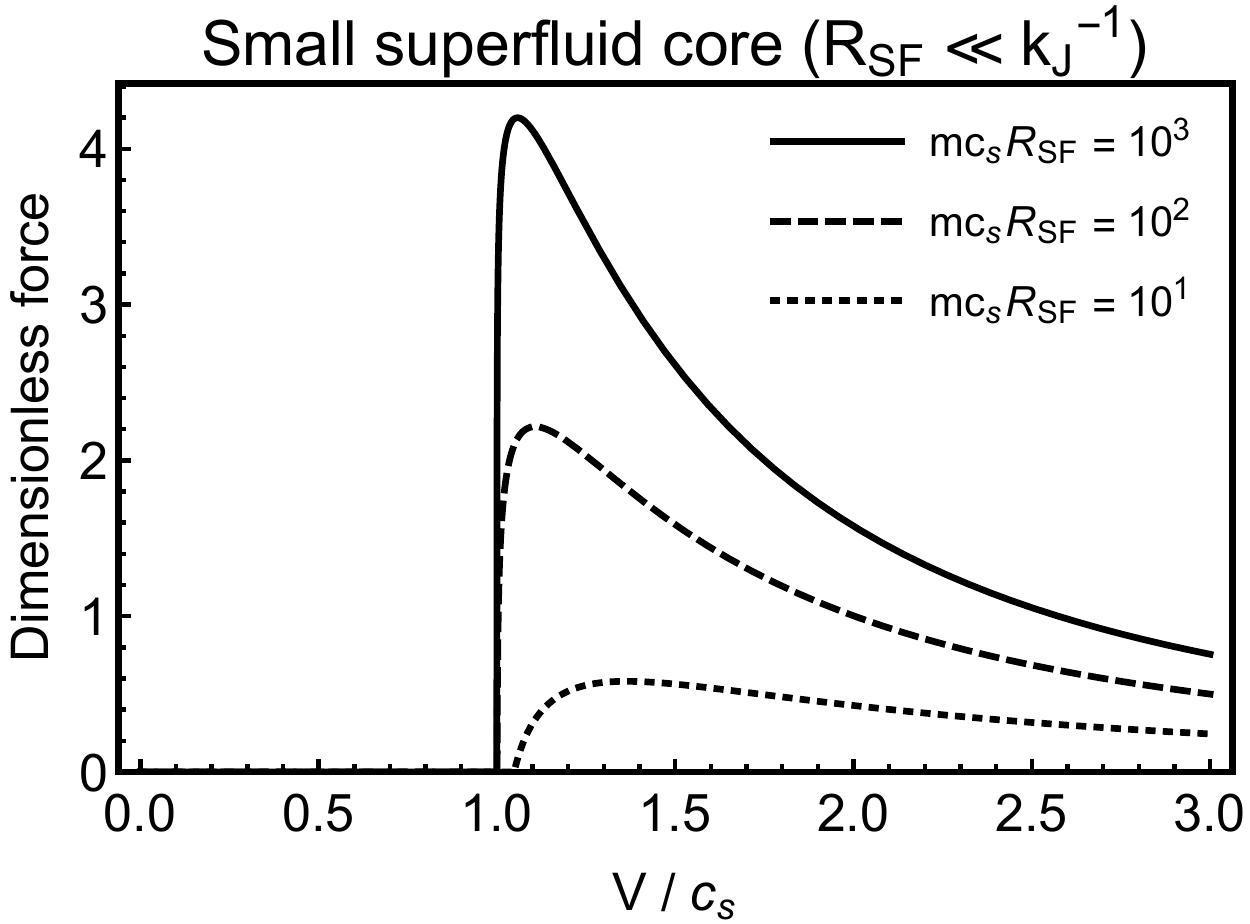}
\includegraphics[width = 0.5\linewidth]{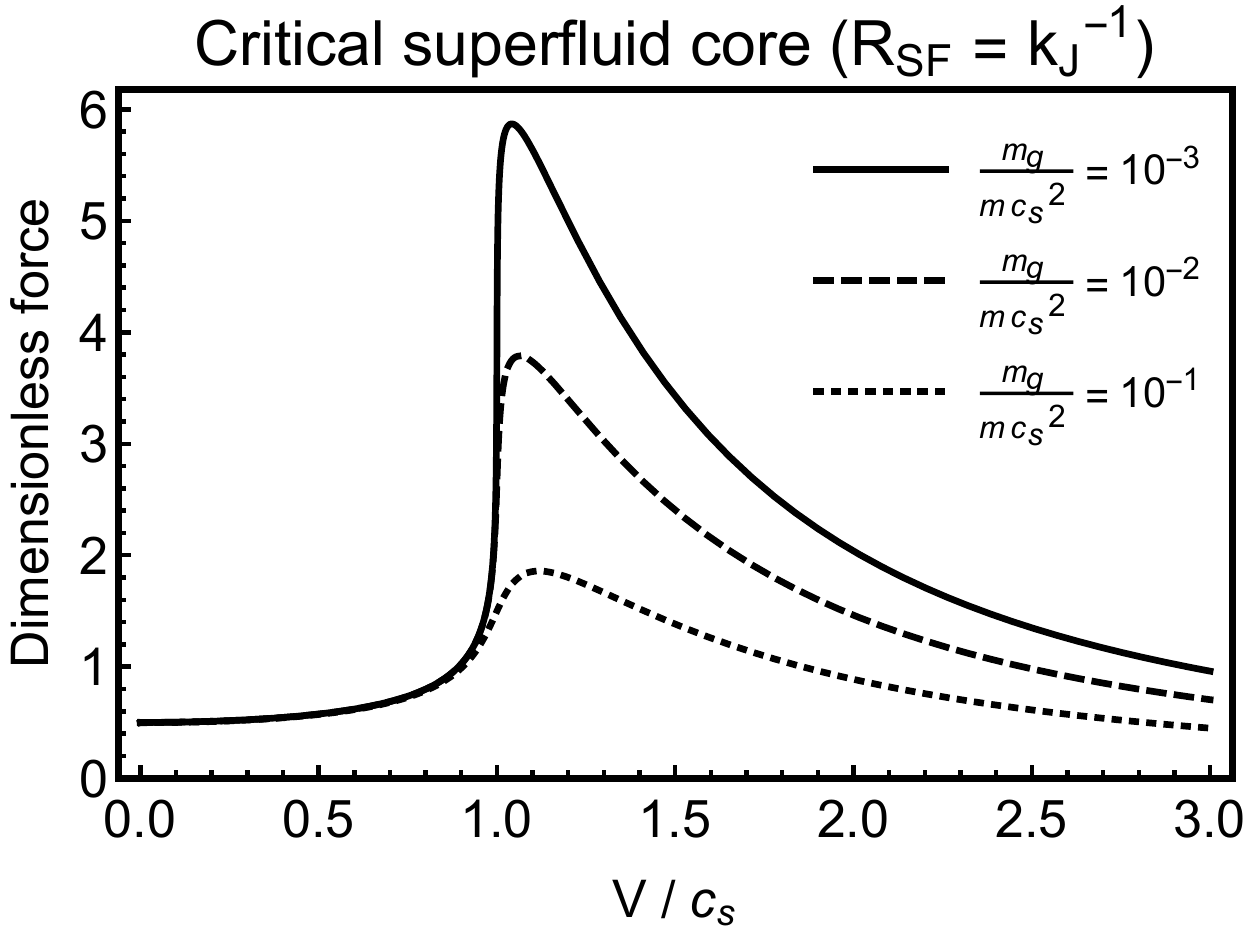}
\caption{\small Dynamical friction force in dimensionless units, $|F| \times \frac{c_s^2}{4\pi G^2M^2\rho_0}$, plotted in two limiting cases. {\it Left:} Friction force~\eqref{DF-qp} valid for small superfluid core, corresponding to $R_{\rm SF} \ll  k_{\rm J}^{-1}$ limit, plotted for various values of $mc_sR_{\rm SF}$. {\it Right:} Friction force~\eqref{F heavy pert} valid for superfluid core of critical size, $R_{\rm SF} =  k_{\rm J}^{-1}$, plotted for various values of $m_g/mc_s^2$.}
\label{DF-plot}
\end{figure}

For simplicity, we consider the case of a sufficiently localized perturber, such that $R_{\rm object}^{-1} \ll k_\star$. It follows from~\eqref{kmax def} that 
the UV cutoff is set by 
\be
k_{\max} = k_\star\,.
\ee
To make headway with calculating $k_\star$, we focus on the case of a heavy perturber, $M\rightarrow\infty$, in which case~\eqref{kstar} admits an analytic solution:
\beq
k_\star^2\simeq 2m^2c_s^2\left( - 1 +  \frac{V^2}{c_s^2} +\sqrt{\left(1- \frac{V^2}{c_s^2}\right)^2+\frac{m_g^2}{m^2c_s^4}} \right)\,.
\label{kstarjeans}
\eeq
One can easily convince oneself that $k_\star > k_{\rm J}$ for arbitrary $V$, hence there is a friction force even for subsonic velocities. In fact, since~\eqref{kstarjeans} coincides with~\eqref{jeansk} for $V=0$, this implies a non-vanishing friction for arbitrarily small (but non-zero) velocities. Indeed, assuming for concreteness that $k_{\max} = k_\star$, the dynamical friction force~\eqref{Fkint} is
\be
|F| = \frac{2\pi G^2M^2\rho_0}{V^2}~{\rm ln}\left(\frac{- 1 +  \frac{V^2}{c_s^2} +\sqrt{\left(1- \frac{V^2}{c_s^2}\right)^2+\frac{m_g^2}{m^2c_s^4}}}{-1+\sqrt{1+\frac{m_g^2}{m^2c_s^4}}}\right)\,.
\label{F heavy pert}
\ee
This is plotted in the Right Panel of Fig.~\ref{DF-plot} for different values of $m_g/mc_s^2$. 

In particular,~\eqref{F heavy pert} greatly simplifies for $V \ll c_s$ and yields a velocity-independent answer:
\be
|F| = \frac{2\pi G^2M^2\rho_0}{c_s^2\sqrt{1+\frac{m_g^2}{m^2c_s^4}}}\,;\qquad V \ll c_s\,.
\label{friction V << c_s}
\ee
The fact that $|F|$ is non-zero for arbitrarily small $V$ may seem puzzling, however one should caution that the results above are valid in the strict $M\rightarrow \infty$ limit. As we are about to show, sub-leading $1/M$ corrections lower the value of $k_\star$, such that $k_\star$ eventually becomes smaller than $k_{\rm J}$ for a small but finite $V$. In other words, the friction force vanishes below a critical velocity $V_{\rm c}$. 

To see this in detail, let us take the limit $V\ll c_s$ from the outset, and solve~\eqref{kstar} perturbatively in $1/M$. The result is
\be
\frac{k_\star}{k_{\rm J}} \simeq 1 + \frac{V}{2c_s^2 \sqrt{1+\frac{m_g^2}{m^2c_s^4}}} \left(V - \frac{k_{\rm J}}{M} \right) + \mathcal{O}\left(\frac{1}{M^2}\right)\,.
\label{kstar pert}
\ee
It is easy to check that this agrees with~\eqref{kstarjeans} for $V\ll c_s$ in the limit $M\rightarrow\infty$. The critical velocity at which $k_\star$ equals $k_{\rm J}$ is readily identified:
\be
V_{\rm c} \equiv \frac{k_{\rm J}}{M}\,.
\label{vcbound}
\ee
In other words, for $V\ll c_s$ the generalization of~\eqref{friction V << c_s} is
\beq
|F|\simeq \frac{2\pi G^2M^2\rho_0}{c_s^2\sqrt{1+\frac{m_g^2}{m^2c_s^4}}}\left(1 - \frac{V_{\rm c}}{V}\right)\,;\qquad V_{\rm c} \leq V \ll c_s\,.
\eeq
Note that $V_{\rm c}$ is small in the limit of a heavy perturber, which is the regime of phenomenological interest.

\section{Accelerated Motion}

Having discussed the free motion of the impurity in some depth, let us switch our attention to accelerated motion. For simplicity we focus on {\it subsonic} accelerated motion
and assume that the superfluid's homogeneity scale is much smaller than the Jeans' scale. The latter amounts to taking $k_{\rm J}\rightarrow 0$, or equivalently $m_g\rightarrow 0$.
The rationale for these simplifications is that the friction force vanishes under these assumptions for uniform ({\it i.e.}, non-accelerated) motion, thereby allowing us to easily pinpoint the effect of acceleration. 

The phenomenon we are after is the superfluid analogue of electromagnetic {\it bremsstrahlung}  radiation. As we know, an accelerating charge emits electromagnetic radiation no matter how small its velocity is. The emitted radiation in turn slows down the charged particle through the so-called ``radiation reaction''. Our goal is to capture a similar effect with phonon emission by a body (or ``impurity'') accelerating through a gravitating superfluid. 

To calculate the phonon radiation reaction self-consistently, one should consider the process of phonon radiation in a scattering process, involving two impurities. The result would be relevant on multiple counts, especially since we are ultimately interested in orbital decay due to phonon emission. However, this calculation would involve evaluating diagrams with more than four external legs --- a straightforward but somewhat involved task. To simplify the calculation, we instead analyze phonon bremsstrahlung emission of a body moving in an {\it external field} of another source. For this purpose we will work with the quasiparticle description.

\subsection{Quasi-particle description}

The relevant Lagrangian is~\eqref{EFT1} in the $m_g\rightarrow 0$ limit:
\beq
\mathcal{L}=\frac{1}{2}\dot{\pi}^2+\frac{1}{2} c_s^2 \pi \Delta \pi
-4\pi G\sqrt{\rho_0}
M^2 \chi^2 \frac{1}{c_s\Delta}\dot{\pi}+\frac{1}{2}\dot{\chi}^2-\frac{1}{2}\big(\vec{\nabla}\chi\big)^2-\frac{1}{2}M^2\chi^2+\delta\mathcal{L}\,.
\label{EFT2}
\eeq
Here $\delta\mathcal{L}$ denotes higher-order terms that are not relevant for tree-level processes with only one external $\pi$-leg attached to external lines,
\beq
\delta \mathcal{L}\equiv-\frac{c_s}{2\sqrt{\rho_0}}\big(\vec{\nabla} \pi\big)^2 \dot{\pi}+\frac{1}{8}\frac{c_s^2}{\rho_0}\big(\vec{\nabla} \pi\big)^4+2\pi GM^2\big(\vec{\nabla} \pi\big)^2 \frac{1}{\Delta} \chi^2-2\pi GM^4\chi^2 \frac{1}{\Delta} \chi^2\,.
\label{EFT2-high}
\eeq
The first two terms originate from the contact interactions among superfluid constituents, whereas the last two interaction terms are of gravitational origin, hence the non-local structure.

In what follows we will assume that the acceleration is caused by some scattering process, the origin of which is unimportant, and compute the rate of radiating phonons from external lines. In general this is not, however, the only channel for radiating phonons when the acceleration is due to gravitational interactions described by the last term of~\eqref{EFT2-high}. Namely, the third term in~\eqref{EFT2} and the second-to-last term in~\eqref{EFT2-high} could also lead to the scattering of impurities mediated by phonon exchange. (These operators originate in~\eqref{withphi} from the mixing of phonons with the Newtonian potential.) For this process, the external phonon line can be attached at more locations besides the external $\chi$-line, such as directly at the interaction vertices or on internal phonon lines. To simplify the calculation we ignore phonon-mediated scattering. This is justified for soft bremsstrahlung, since the operator  $\big(\vec{\nabla} \pi\big)^2 \Delta^{-1} \chi^2$ (and others like it) involves more derivatives than the operator $\chi^2 \Delta^{-1} \chi^2$ of interest.

\begin{figure}[t]
\centering
\includegraphics[width=5in]{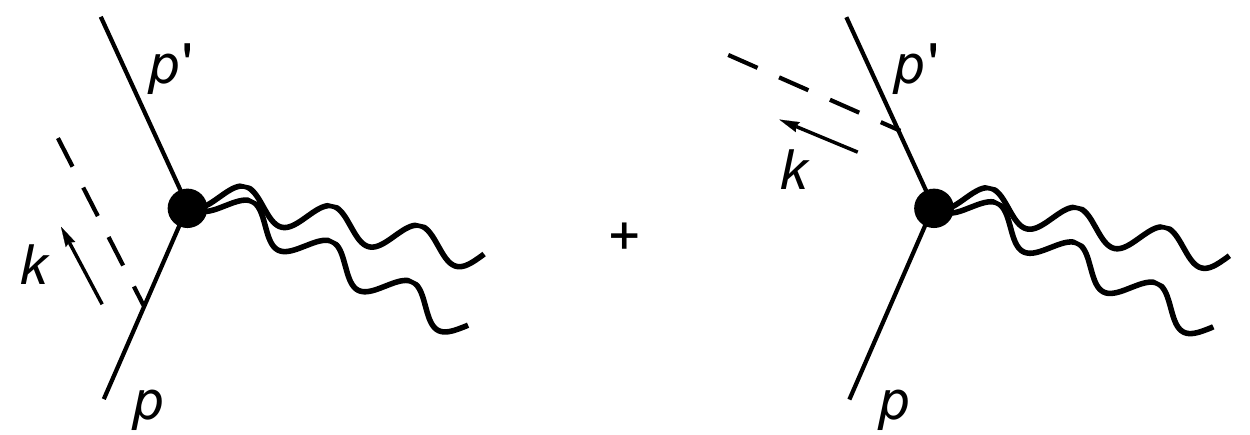}
\caption{\label{brem} The relevant diagrams for soft phonon bremsstrahlung. Wiggly lines denote the external force causing the acceleration of the incoming particle, the solid line denotes the impurity, and the dashed line is the radiated phonon.}
\end{figure}

Therefore, following the bremsstrahlung calculation from~\cite{Peskin:1995ev} we are interested in the diagrams shown in Fig.~\ref{brem}. Let $p$ and $p'$ denote the initial and final four-momenta of $\chi$, respectively, while $k$ is the four-momentum of the radiated phonon. From the effective Lagrangian we can immediately infer the following propagators and vertices:

\begin{itemize}

\item \textit{Phonon propagator:}~~~~\; $D^{\pi}(\omega_k,\vec{k})=\frac{-{\rm i}}{-\omega_k^2+c_s^2\vec{k}^2}$.

\item \textit{Impurity propagator:}~~~ $D^{\chi}(p)=\frac{-{\rm i}}{p^2+M^2}$.

\item \textit{Interaction vertex:}~~~~~~\; $\Gamma^{\pi\pi\pi}(k_1,k_2,k_3)=\frac{c_s}{\sqrt{\rho_0}}\left( (\vec{k}_1\cdot \vec{k}_2) \omega_{k_3}+(\vec{k}_1\cdot \vec{k}_3) \omega_{k_2}+(\vec{k}_2\cdot \vec{k}_3) \omega_{k_1} \right)$.

\item \textit{Interaction vertex:}~~~~~~\; $\Gamma^{\chi\chi\pi}(p_1,p_2,k)=\frac{4\pi GM^2\sqrt{\rho_0}}{c_s}\frac{\omega_k}{\vec{k}^2}$.

\item \textit{Interaction vertex:}~~~~~~~ $\Gamma^{\chi\chi\pi\pi}(p_1,p_2,k_1,k_2)=-4\pi {\rm i}GM^2 \frac{\vec{k}_1\cdot \vec{k}_2}{(\vec{k}_1+\vec{k}_2)^2}$.
\end{itemize}

\vspace{0.2cm}
\noindent Using the above Feynman rules the amplitude for the diagrams in Fig.~\ref{brem} follows readily:  
\beq
\mathcal{A}\propto \frac{G M^2\sqrt{\rho_0}}{c_s}\frac{\omega_k}{\vec{k}^2}\left( \frac{\mathcal{A}_0(p-k,p')}{(p-k)^2+M^2}+  \frac{\mathcal{A}_0(p,p'+k)}{(p'+k)^2+M^2}\right)\,,
\eeq
where we ignore numerical coefficients for simplicity, focusing instead on the parametric dependence. Here, $\mathcal{A}_0$ denotes the amplitude without external phonon leg.  
For soft phonon emission, $|\vec{k} |\ll |\vec{p}\,'-\vec{p}|$, we can approximate the latter by $\mathcal{A}_0(p,p'+k)\approx  \mathcal{A}_0(p-k,p')\approx  \mathcal{A}_0(p,p')$, hence
the amplitude at leading order in soft phonon momentum reduces to
\beq
\mathcal{A}\propto\mathcal{A}_0(p,p') \frac{GM\sqrt{\rho_0}}{c_s} \frac{\omega_k}{\vec{k}^2} \left( \frac{1}{p\cdot k/M}-\frac{1}{p'\cdot k/M} \right)\,.
\label{AA0}
\eeq

The differential probability of radiating a phonon with momentum $\vec{k}$ can be expressed in terms of the amplitudes as
\beq
{\rm d}{\cal P}\propto \frac{{\rm d}^3 k}{\omega_k} \left| \frac{\mathcal{A}}{\mathcal{A}_0} \right|^2 \,.
\eeq
We can substitute~\eqref{AA0} and use the on-shell dispersion relation $\omega_k = c_sk$, where $k = |\vec{k}|$ henceforth denotes the magnitude
of the 3-momentum. Expressing the result in terms of the initial velocity $V$ and final velocity $V'$ of the perturber $\chi$, we obtain
\ba
\nonumber
{\rm d}{\cal P} &\propto & \frac{G^2M^2\rho_0}{c_s}\frac{{\rm d}^3 k}{k^3}\, \left( \frac{1}{\left(c_sk-\vec{k}\cdot \vec{V}\right)^2}+\frac{1}{\left(c_sk-\vec{k}\cdot \vec{V'}\right)^2}-\frac{2}{\left(c_sk-\vec{k}\cdot \vec{V}\right)\left(c_sk-\vec{k}\cdot \vec{V}'\right)}\right)\\
& \simeq & G^2M^2\rho_0 \frac{{\rm d}^3 k}{c_s^5 k^5} \left(\hat{k}\cdot \big(\vec{V}-\vec{V}'\big)\right)^2\,,
\ea
where in the last step we have taken the subsonic limit $V\ll c_s$.  Adopting spherical coordinates and performing the angular integral, we are left with 
\beq
{\rm d}{\cal P}\propto G^2M^2\rho_0 \big( \vec{V}-\vec{V'}\big)^2 \frac{{\rm d}k}{c_s^5k^3} \,.
\eeq
Therefore, the total energy radiated in phonons, as the perturber velocity changes from $V$ to $V'$ due to an external field, is 
\be
E_{\rm rad} = \int {\rm d}k \, \omega_k\frac{{\rm d}{\cal P}}{{\rm d}k} \propto \frac{G^2M^2\rho_0}{c_s^4} \big( \vec{V}-\vec{V'}\big)^2 \int_{k_{\rm min}}^{k_{\rm max}} \frac{{\rm d}k}{k^2}\,.
\label{Erad quantum}
\ee
The lower bound of integration is given by size of the superfluid core, $k_{\rm min}  = 2\pi R_{\rm SF}^{-1}$, while the upper bound is given by the momentum at which phonons stop behaving like waves, namely $k_{\rm max}=2mc_s$. 

Equation~\eqref{Erad quantum} displays some expected features. The dependence on $G$, $M$ and $\rho_0$ is exactly the same as
in our earlier expression for a freely-moving perturber. The dependence on $\big(\vec{V}-\vec{V}'\big)^2$ is also expected, since similar radiation
for accelerated electric charges depends on the square of the acceleration. A surprising aspect of $E_{\rm rad}$ is its IR divergent behavior. Instead of a
logarithmic dependence, as in the free case, the radiated energy is power-law divergent:
\beq
E_{\rm rad}  \sim  G^2M^2\rho_0 \frac{\big( \vec{V}-\vec{V'}\big)^2}{c_s^4 k_{\rm min}} \sim G^2M^2\rho_0 \frac{\big( \vec{V}-\vec{V'}\big)^2R_{\rm SF} }{c_s^4}\,.
\label{qm-scatt}
\eeq
This traces back to the non-local form of the operator $\chi^2 \Delta^{-1} \chi^2$.

It is instructive to contrast~\eqref{qm-scatt} with the analogous result for bremsstrahlung radiation in electrodynamics:
\beq
{\rm d}E = {\rm d}k ~e^2\left(\vec{V}-\vec{V}'\right)^2\,.
\eeq
This looks similar to the integrand in~\eqref{Erad quantum}, except that the result is finite in the IR and instead diverges in the UV. The difference of course originates
from the photon-charged particle interaction being local. Since the photon momentum is assumed soft, it makes sense to argue that it must be cutoff at the time scale
of the kick that led to $V\rightarrow V'$. With this regulator, we recover the well-known expression for the rate of radiated energy,
\beq
\frac{\Delta E}{\Delta t}= e^2 \left(\frac{\Delta V}{\Delta t}\right)^2\,,
\eeq
where $\frac{\Delta V}{\Delta t}$ is recognized as the acceleration.

\subsection{Classical computation}

It is instructive to rederive the energy radiated into phonons using a classical bremsstrahlung computation. The starting point is the classical phonon equation of motion
\beq
\ddot{\pi}-c_s^2 \Delta \pi=\frac{G\sqrt{\rho_0}}{c_s}\frac{\partial_t}{\Delta}\rho_{\chi}\equiv J\,.
\label{weq}
\eeq
The source $J$ is given by the non-relativistic energy density $\rho_\chi$ of the $\chi$ field:
\be
J(x) \equiv  \frac{G\sqrt{\rho_0}}{c_s}\frac{\partial_t}{\Delta}\rho_{\chi}\,.
\ee
For a particle moving along a trajectory $y^\mu(\tau)$, with proper time $\tau$, 
\beq
\rho_\chi = M \int {\rm d}\tau\, \delta^{(4)}\big(x^\mu-y^\mu(\tau)\big)\,.
\eeq

We are interested in the process of a sudden kick. In other words, the $\chi$ particle is considered to be moving with momentum $p$ until $\tau=0$, and then instantaneously starts moving with momentum $p'$. (Once more we are closely following the analogous calculation for electrodynamics from~\cite{Peskin:1995ev}.) In this case the current sourcing the wave equation~\eqref{weq} takes the following form
\beq
J(x)=\frac{G\sqrt{\rho_0}M}{c_s}\frac{\partial_t}{\Delta}\left\{ \int_0^\infty {\rm d}\tau\, \delta^{(4)}\left(x^\mu-\frac{p'^{\mu}}{M}\tau\right)+\int_{-\infty}^0 {\rm d}\tau \,\delta^{(4)}\left(x^\mu-\frac{p^\mu}{M}\tau\right) \right\} \,.
\eeq
Its Fourier transform is
\beq
\tilde{J}(k)\propto \frac{G\sqrt{\rho_0}M}{c_s}\frac{k_0}{\vec{k}^2}\left( \frac{1}{k_\mu p'^\mu /M+i\epsilon}-\frac{1}{k_\mu p^\mu/M-i\epsilon} \right)\,,
\eeq
where the ${\rm i}\epsilon$ regulators ensure convergence of the integrals. This current in turn leads to the following solution to~\eqref{weq}, for the radiation configuration,
\beq
\pi_{\rm rad}(x)\propto\frac{G\sqrt{\rho_0}M^2}{c_s} \int  \frac{{\rm d}^3k}{\vec{k}^2}\left\{ {\rm e}^{{\rm i} k\cdot x} \left( \frac{1}{k_\mu p'^\mu}-\frac{1}{k_\mu p^\mu} \right)+{\rm c.c.} \right\}\,.
\eeq

Assuming that nonlinearities are negligible, this can be converted into radiation energy using the free phonon Hamiltonian
\beq
E_{\rm rad}=\int {\rm d}^3x \left( \frac{1}{2}\dot{\pi}_{\rm rad}^2+\frac{1}{2}c_s^2(\vec{\nabla}\pi_{\rm rad})^2 \right)\,.
\eeq
After substituting the radiation configuration and simplifying, we get
\beq
E_{\rm rad}\propto G^2M^2\rho_0  \int {\rm d}{\rm cos}\theta ~{\rm d}k \left.\left( \frac{1}{k_\mu p'^\mu}-\frac{1}{k_\mu p^\mu} \right)^2\right\vert_{k_0=c_s k}\sin^2\left( c_skt \right)\,.
\eeq
In the subsonic limit $V\ll c_s$ the expression simplifies further to give
\beq
E_{\rm rad}\propto \frac{G^2M^2\rho_0}{c_s^4}(\vec{V}-\vec{V'})^2\int {\rm d}k \,\frac{{\rm sin}^2\left( c_skt \right)}{k^2}\,.
\eeq
After averaging this expression over times longer than $k_{\rm min}^{-1}$, and integrating it from $k_{\rm min}$ to $k_{\rm max}\gg k_{\rm min}$, we arrive at the expression identical to the result of the quantum scattering~\eqref{qm-scatt}.
Notice that this method can be straightforwardly applied to other types of motion, {\it e.g.}, uniformly-accelerated motion.

\section{Conclusions}

The phenomenon of dynamical friction in gravitating systems is a venerable subject dating back to Chandrasekhar's seminal work. In this paper we have calculated the dynamical friction experienced by a perturber moving through an inviscid fluid medium, {\it i.e.}, a superfluid. Crucially, the calculation accounts for both the tachyonic mass term responsible for the Jeans instability and higher-gradient corrections in the dispersion relation, with the latter being responsible for ``quantum pressure'' in the hydrodynamical language. We have performed the calculation in two ways, first using a quasiparticle description of phonon radiation, and second with a standard linearized treatment of the hydrodynamical equations. We considered the case of a perturber moving at uniform velocity, as well as that of an accelerated perturber due to an external field.

Our calculation resolves a puzzling aspect of the standard steady-state result~\cite{Dokuchaev:1964,Ruderman:1971,Rephaeli:1980,Ostriker:1998fa}, given by~\eqref{DF-nomass}, namely the apparent discontinuity as $V\rightarrow c_s$. We have shown that including either the Jeans tachyonic mass and/or the higher-gradient correction in the phonon dispersion relation is sufficient to make the dynamical friction force a continuous function of velocity. Furthermore, we found that the Jeans instability can result in a non-zero dynamical friction force even for subsonic motion.

Strictly speaking, the application of~\eqref{highderdisp} to phonons with wavelength comparable to the Jeans scale, in case of a localized superfluid soliton, is merely an intelligent guess, as underscored in the Introduction. For a proper account of such modes a more detailed analysis of the soliton's spectrum is required. It would be particularly interesting to see whether the precise spectrum is continuous or discrete, and to study the subsequent impact on the drag force.

Furthermore, in this work we have considered the energy dissipation for limited types of motion. It would be interesting to investigate more realistic processes, such as orbiting and spinning sources of different shapes. The best way to tackle this problem may be to evaluate the classical phonon radiation field at large distances, which would simplify the computation significantly.

As mentioned from the outset, our interest in dynamical friction stems in part from recent developments in dark matter model building, in particular superfluid~\cite{Goodman:2000tg,Berezhiani:2015pia,Berezhiani:2015bqa,Berezhiani:2017tth} and fuzzy~\cite{Hu:2000ke,Hui:2016ltb} dark matter models. In both scenarios, dark matter undergoes Bose-Einstein condensation in the central regions of galactic halos.

As an application of the results derived in this paper, it will be interesting to quantify whether superfluid dark matter can alleviate some of the minor problems of standard cold dark matter related to dynamical friction. For instance, we expect that galactic bars in spiral galaxies should achieve a nearly constant velocity, as favored by observations~\cite{Debattista:1997bi}, instead of being slowed down by dynamical friction. As argued in fuzzy dark matter~\cite{Hui:2016ltb}, the suppression of dynamical friction can also offer a natural explanation to the long-standing puzzle of why the five globular clusters orbiting Fornax have not merged to the center to form a stellar nucleus~\cite{Fornaxold,Tremaine}. Finally, it will be interesting to estimate the impact on structure formation, in particular on the rate of galaxy mergers. One would expect that the suppression of dynamical friction for subsonic mergers should result on average in longer merger times involving multiple encounters.

\section*{Acknowledgments}
The authors are grateful to Clare Burrage, Gia Dvali, Lam Hui, Austin Joyce, Peter Millington, Alberto Nicolis, Georg Raffelt, and Rachel Rosen for helpful discussions, and especially to Yoni Kahn and Mariangela Lisanti. B.E. is supported by a Leverhulme Trust Research Leadership Award. J.K. is supported in part by the US Department of Energy (HEP) Award DE-SC0013528, NASA ATP grant 80NSSC18K0694, the Charles E. Kaufman Foundation of the Pittsburgh Foundation, and a W.~M.~Keck Foundation Science and Engineering Grant.

\end{document}